\documentclass[]{jfm}

\usepackage{graphicx}
\usepackage{newtxtext}
\usepackage{newtxmath}
\usepackage{amsmath,amssymb}
\usepackage{bm}
\usepackage{xcolor}
\usepackage{natbib}
\usepackage{hyperref}
\hypersetup{colorlinks=true,urlcolor=blue,citecolor=black,linkcolor=black}
\usepackage{soul}

\title{Evolution of inertial particle clustering in decaying turbulence}

\author{Mart\'in Obligado\aff{1,2}
  \and Sof\'ia Angriman\aff{3}\corresp{\email{sofia.angriman@univ-cotedazur.fr}}}

\affiliation{\aff{1}Univ. Lille, CNRS, ONERA, Arts et M\'etiers Institute of Technology,
Centrale Lille, UMR 9014--LMFL--Laboratoire de M\'ecanique des Fluides de Lille--Kamp\'e de F\'eriet,
F-59000 Lille, France\\[\affilskip]
\aff{2}Institut universitaire de France (IUF), Paris, France\\[\affilskip]
\aff{3}Universit\'e C\^ote d'Azur, CNRS, Institut de Physique de Nice (INPHYNI),
17 rue Julien Laupr\^etre, 06200 Nice, France}

\begin{document}

\maketitle

\begin{abstract}
Inertial particle dynamics in turbulence are commonly interpreted in terms of the instantaneous particle Stokes number, an approach supported by extensive studies in statistically stationary homogeneous isotropic turbulence. In freely decaying turbulence, however, the Kolmogorov time scale evolves continuously, causing the effective Stokes number to vary even though particle properties remain unchanged. Whether preferential concentration remains uniquely determined by this instantaneous Stokes number or depends on the flow's previous evolution has remained largely unexplored.

We investigate inertial-particle clustering during the decay of homogeneous isotropic turbulence using direct numerical simulations of heavy point particles in the one-way coupling limit. Clustering is quantified through three-dimensional Vorono\"i tessellations for populations evolving either from statistically stationary clustered states or initially random distributions. We examine the influence of the turbulent initial condition by considering both conventionally forced homogeneous isotropic turbulence and velocity fields reconstructed using physics-informed neural networks.

We show that preferential concentration cannot be described solely by the instantaneous Stokes number. While particle slip velocity follows the classical steady-state dependence on the instantaneous Stokes number, clustering statistics retain a measurable history dependence throughout decay. Existing stationary scaling laws remain applicable to cluster-size evolution, but only through history-dependent prefactors. Particle distributions evolving from different initial conditions rapidly converge towards similar large-scale spatial organisation, indicating that the carrier flow determines the geometry of preferential concentration while finer clustering statistics preserve imprints of the previous evolution. These results demonstrate that preferential concentration in non-stationary turbulence is governed jointly by the instantaneous flow state and its initial state.

\end{abstract}

\section{Introduction}

The transport of inertial particles by turbulent flows is a fundamental problem in fluid mechanics, governing a wide range of environmental and industrial processes. In the atmosphere, turbulence controls the dispersion and clustering of cloud droplets, influencing collision and coalescence rates and, consequently, aspects of precipitation formation. Similar mechanisms arise in river and estuarine plumes, volcanic ash transport, marine ecosystems, sediment transport, combustion systems, sprays, and numerous multiphase engineering applications. In all these situations, the interaction between particles and turbulence determines the efficiency of transport, mixing, and reaction processes.

Unlike fluid tracers, inertial particles cannot instantaneously adjust their velocity to that of the surrounding flow. Instead, their finite response time leads to preferential sampling of the turbulent field, producing highly inhomogeneous spatial distributions known as preferential concentration. Since the pioneering work of \citet{Maxey1987}, preferential concentration has been recognised as one of the defining features of inertial particle dynamics. Numerous theoretical, numerical, and experimental studies have demonstrated that turbulence induces the formation of intense particle clusters through inertial effects, producing concentration fluctuations that may exceed an order of magnitude relative to a random distribution~\citep{Falkovich2002,Bec2003,Monchaux2010,Sumbekova2017}.

Most studies of inertial particle dynamics have focused on statistically
stationary homogeneous isotropic turbulence, where the particle Stokes number
remains constant \citep{liu2020life}. Under these conditions, preferential concentration
\citep{Bec2003,Toschi2009}, relative particle velocities, and collision rates
\citep{Gustavsson2016,brandt2022particle} are now relatively well understood.
Freely decaying homogeneous isotropic turbulence, on the other hand, has served
as a canonical problem in turbulence research since the pioneering work of
\citet{vonKarman1948}. Beyond its fundamental importance, it also provides an
ideal model for many natural and engineering flows in which turbulence evolves
after the forcing has ceased, such as atmospheric boundary layers, river
plumes, volcanic clouds, and industrial mixing devices. The decay of the
single-phase turbulent field has therefore been extensively investigated,
including its energy decay, self-similarity and spectral evolution
\citep{George1992,Sreenivasan1998,Davidson2015}. Comparatively little
attention, however, has been devoted to the evolution of inertial particles
under these conditions \citep{Sarkar2025}.

The distinction between stationary and decaying turbulence is particularly important because the particle Stokes number is no longer constant. Preferential concentration is commonly characterised, for sub-Kolmogorov dense particles, through the Stokes number,
\begin{equation}
\mathrm{St}=\frac{\tau_p}{\tau_\eta},
\label{eq:stokes}
\end{equation}
where $\tau_p$ is the particle response time and $\tau_\eta=\sqrt{\nu/\varepsilon}$ is the Kolmogorov time scale, with $\varepsilon=\nu\langle\omega^2\rangle$ the energy dissipation rate, $\nu$ the kinematic viscosity, and $\langle\omega^2\rangle$ the mean enstrophy. During turbulence decay, $\tau_\eta$ continuously increases, causing the effective particle Stokes number to evolve even though the particle properties remain fixed.

The dependence of preferential concentration on the Stokes number has been extensively documented for statistically stationary turbulence
\citep{Eaton1994,Toschi2009,brandt2022particle}. Preferential concentration exhibits a non-monotonic dependence on particle inertia, with clustering intensity increasing from the tracer limit to a maximum near $\mathrm{St}=\mathcal{O}(1)$ \citep{Bec2003,Falkovich2002} before decreasing for larger Stokes numbers as particles progressively decouple from the dissipative scales \citep{Gustavsson2016,brandt2022particle}. Although several physical mechanisms have been proposed to explain this behaviour, including centrifugal ejection \citep{Maxey1987}, sweep--stick dynamics \citep{Goto2008,Falkovich2002}, and non-local path-history effects \citep{Bragg2015JFM}, they all distinguish between particles with $\mathrm{St}\lesssim1$ and $\mathrm{St}\gtrsim1$, where different physical processes dominate the clustering dynamics. Decaying turbulence fundamentally alters this picture: particles can potentially evolve across the $\mathrm{St}\approx1$ transition while retaining the imprint of their previous interaction with the turbulent flow. Consequently, it remains unclear whether the instantaneous Stokes number is sufficient to determine preferential concentration during turbulence decay, or whether the clustering dynamics also depend on the preceding evolution of the system.

The objective of the present work is to investigate the temporal evolution of inertial-particle clustering during the decay of homogeneous isotropic turbulence (HIT), with particular emphasis on the role of the initial conditions. Rather than considering only particles introduced randomly at the onset of the decay, we also examine particle distributions that have already developed preferential concentration under statistically stationary conditions before the external forcing is removed. This allows us to distinguish between the response of an initially random particle field and the persistence and reorganisation of pre-existing clusters as the turbulence decays. By comparing randomly distributed and clustered particles evolving within the same turbulent flow, we directly assess whether the subsequent clustering dynamics are determined solely by the instantaneous flow properties or if they retain a measurable dependence on the initial particle organisation.

This distinction is important because the particle distribution at the beginning of the decay is not merely a passive initial condition. Pre-existing clusters are characterised by reduced inter-particle distances, strong concentration fluctuations and a broad distribution of local particle densities. Their subsequent evolution may therefore differ substantially from that of particles initially distributed at random, even when both populations are subjected to the same velocity field and have the same particle response time. The comparison provides a direct means of quantifying the memory associated with the clustering itself, independently of changes in the carrier flow. To the best of our knowledge, the influence of an initially clustered particle distribution on the subsequent evolution of preferential concentration in decaying turbulence has not been systematically investigated.

Direct numerical simulations are performed at moderate initial Taylor-scale Reynolds numbers ($\mathit{Re}_\lambda = \sqrt{15u^4/\nu\varepsilon}\approx200$, with $u$ the root-mean-square velocity), allowing the entire decay process to be resolved while maintaining sufficiently developed small-scale turbulence to generate intense preferential concentration. Since the primary objective of this work is to investigate the evolution of particle clustering, we consider the canonical configuration of heavy point particles (with diameters smaller than the Kolmogorov length scale) in the absence of gravity and in the one-way coupling limit. This simplified framework isolates the inertial mechanisms responsible for preferential concentration while enabling direct comparison with the theoretical models developed for statistically stationary turbulence. Particle clustering is quantified using Vorono\"i tessellations, which provide a robust framework for measuring clustering intensity, identifying coherent particle clusters and characterising their geometrical properties~\citep{Monchaux2010,Sumbekova2017}.

In addition to the comparison between randomly initialised and pre-clustered particles in conventionally forced HIT, we investigate the sensitivity of the particle dynamics to the initial turbulent organisation. Besides classical turbulence sustained through random large-scale forcing, we consider turbulent initial conditions generated using Physics-Informed Neural Networks (PINNs) combined with statistical data assimilation~\citep{Angriman2023,Angriman2024}. These reconstructed velocity fields preserve homogeneous isotropic conditions while reproducing selected statistical features observed in active-grid-generated turbulence. Together, these configurations allow us to assess separately the influence of the initial particle distribution and that of the initial turbulent organisation on the subsequent decay.

The present study therefore addresses two complementary forms of dependence on the initial conditions. The first is associated with the initial spatial organisation of the particles and is isolated by comparing randomly distributed and pre-clustered particles subjected to the same decaying HIT. The second is associated with the initial organisation and forcing history of the turbulent flow and is examined by comparing clustered particle populations generated under different statistically stationary conditions. This framework makes it possible to determine whether preferential concentration during decay is governed primarily by the instantaneous Stokes number and turbulence intensity, or whether it also depends on the previous evolution of both the carrier flow and the particle distribution. Beyond its fundamental relevance to non-equilibrium particle-laden turbulence, this problem is of applied interest because particles in natural and industrial flows are rarely introduced into a perfectly homogeneous distribution and frequently enter decaying or transient turbulent regions after clustering has already developed.

We show that the evolution of preferential concentration cannot be interpreted solely in terms of the instantaneous Stokes number. Instead, particle clustering remains history dependent long after the onset of turbulence decay, retaining a strong dependence on the initial Stokes number. These findings provide new insight into inertial-particle dynamics under non-stationary conditions and have direct implications for particle transport in naturally evolving turbulent flows.

The manuscript is organised as follows. Section~\ref{sec:numerical} presents the numerical methodology and describes the generation of the different turbulent initial conditions. Section~\ref{sec:steady} investigates the evolution of particle clusters originating from statistically stationary turbulence. Section~\ref{sec:initial} examines the influence of the initial turbulent state on the persistence of preferential concentration and discusses the role of memory effects. Finally, Section~\ref{sec:discussion} summarises the main conclusions and outlines the implications of the present findings for environmental and industrial particle-laden flows.

\section{Numerical setup}\label{sec:numerical}
We consider the free decay of turbulent states laden with inertial particles. For this, we solve the incompressible Navier-Stokes equations
\begin{equation}
    \frac{\partial}{\partial t} {\bf u} + ({\bf u} \cdot {\bm \nabla}) {\bf u} = -\frac{1}{\rho_\mathrm{f}} {\bm \nabla} p + \nu \nabla^2 {\bf u},
    \label{eq:NS}
\end{equation}
where ${\bf u}$ is the incompressible (${\bf \nabla} \cdot {\bf u} = 0$) velocity field, $\rho_\mathrm{f}$ is the fluid density, and $\nu$ is the kinematic viscosity. The equations are written in units of a reference length $L^\mathrm{DNS}$ and reference velocity $U^\mathrm{DNS}$, and solved in a box of size $2\pi L^\mathrm{DNS}$ with periodic boundary conditions, using the parallel pseudo-spectral code GHOST~\citep{Gomez2005,Mininni2011}. We use a uniform mesh in all directions, with $N^3 = 512^3$ grid points.

Along with the flow, we evolve $N_p = 10^6$ inertial point particles subjected to Stokes drag, whose equation of motion is given by
\begin{equation}
\dot{\mathbf{x}}_\text{p} = \mathbf{v}(t), \quad \dot{\mathbf{v}} = \frac{1}{\tau_p}[\mathbf{u}(\mathbf{x}_\text{p},t) - \mathbf{v}(t)].
\end{equation}
Here, $\mathbf{x}_\text{p}$ and $\mathbf{v}$ are the particle position and velocity, respectively, and $\tau_p$ is the particle response time. The particles are one-way coupled to flow, i.e. they do not interact with each other and do not exert a feedback on the flow. As discussed in the previous section, the inertia of the particles is quantified by the Stokes number $\mathrm{St}$. 

Equations~\eqref{eq:NS} are first solved in the presence of a volumetric random forcing $\bm{f}$ that sustains statistically stationary homogeneous isotropic turbulence. Energy is injected into the Fourier shell $kL^\mathrm{DNS}\in[1,3]$ with fixed amplitude and slowly varying phases, characterised by a correlation time of $0.05\,L^\mathrm{DNS}/U^\mathrm{DNS}$. Once statistical stationarity has been reached, inertial particles are introduced at random positions and evolved in the forced flow until their clustering statistics reach a statistically stationary state. The forcing is then removed and the turbulence is allowed to decay freely. These pre-clustered particle populations constitute the reference configuration for the decay. To span a broad range of particle inertia and clustering regimes, initial Stokes numbers $\mathrm{St}(t=0)=0.5$, 3, 6 and 10 are considered. In addition, for $\mathrm{St}(t=0)=6$, we consider a population that is randomly distributed at the instant the forcing is removed. Comparing the pre-clustered and randomly initialised populations evolving in the same decaying velocity field allows us to isolate the effect of the initial particle organisation and, in particular, to determine whether preferential concentration is governed solely by the instantaneous flow state or retains a measurable dependence on the preceding particle dynamics.

To assess independently the role of the initial turbulent organisation, we also consider two classes of freely decaying flows. The first corresponds to conventional HIT generated by the large-scale random forcing described above. The second is initialised from velocity fields reconstructed using physics-informed neural networks (PINNs) combined with the data-assimilation technique known as \textit{nudging}. As shown by \citet{Angriman2024}, these reconstructed fields retain the large-scale and second-order characteristics of homogeneous isotropic turbulence while exhibiting systematic departures in their third-order statistics, consistent with anomalies observed experimentally in active-grid-generated turbulence. Details of the reconstruction procedure are provided in appendix~\ref{sec:methods}. The purpose here is not to investigate the origin of these statistical differences, but rather to exploit the reconstructed fields as distinct turbulent initial conditions with comparable second-order properties but different spatial organisations and decay histories. Comparing their subsequent evolution therefore provides a controlled means of assessing the sensitivity of preferential concentration to the state and history of the carrier turbulence. For this comparison, particles with $\mathrm{St}(t=0)=6$ are used in both flows.

\begin{figure}
\centering
\includegraphics[width=\textwidth]{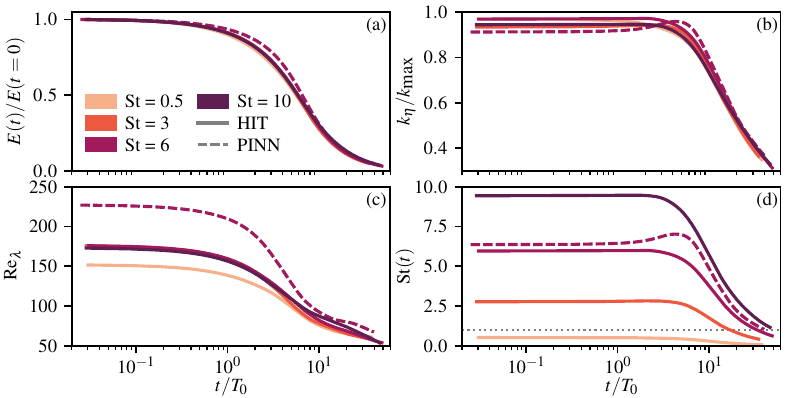}
\hfill
  \caption{Time evolution of global quantities. (a) Flow kinetic energy, normalised by the energy at $t=0$. (b) Wave-number associated to the smallest dynamical scales of the flow, normalised by the maximum wave-number resolved in the simulations $k_\text{max}$. (c) Taylor-scale Reynolds number. (d) Time-dependent Stokes number $\mathrm{St}(t) = \tau_p/\tau_\eta(t)$. The dotted horizontal line corresponds to $\mathrm{St} = 1$. The labels for all panels are the same as in panel (a). The different colours correspond to initial values of the Stokes number, and the line style indicates the flow type (either HIT or PINN).}
\label{fig:flow_properties}
\end{figure}

The temporal evolution of several global flow quantities is presented in figure~\ref{fig:flow_properties}. Time is normalised by the integral time-scale $T_0$ at $t=0$, defined as $T_0 = L_0/U_0$, with $U_0$ the root mean square value of the flow velocity, and $L_0$ the integral length-scale computed from the kinetic energy spectrum $E(k)$,
\begin{equation}
L_0 = \frac{\pi}{4}\frac{\int E(k) k^{-1} dk}{\int E(k) dk}.
\end{equation}
Typical values of $L_0$ are of the order of $L_0\approx L^\text{DNS}/4$.
 As previously reported by \citet{Angriman2024}, the two turbulent fields (HIT and PINN) exhibit different decay dynamics (figure~\ref{fig:flow_properties}(a)). In particular, the PINN reconstruction displays a slower decay until entering the classical decay regime. We stress, however, that the objective of the present work is not to investigate the universality of turbulence decay, but rather to exploit these different decay histories as distinct environments in which inertial particles evolve.

Figure~\ref{fig:flow_properties}b demonstrates that the smallest dynamically relevant scales remain adequately resolved throughout the simulations. The ratio between the Kolmogorov wavenumber and the maximum resolved wavenumber remains below unity throughout the decay, confirming adequate resolution of the dissipative scales. Simultaneously, the Taylor-scale Reynolds number (figure~\ref{fig:flow_properties}c) decreases monotonically from approximately $\text{Re}_\lambda \approx 150$--$230$, depending on the initial condition, down to values close to $50$, covering a broad range of turbulence states while remaining sufficiently turbulent to sustain inertial particle dynamics. We also remark that, during most of the decay, the PINN DNS exhibits significantly larger values of $\text{Re}_\lambda$.

A central quantity for the present study is the instantaneous Stokes number, shown in figure~\ref{fig:flow_properties}d. Although the particle response time remains constant throughout the simulations, the Kolmogorov time scale evolves continuously as the turbulence decays, causing the effective Stokes number to change in time. For the HIT simulations, this evolution is essentially monotonic, with particles with initial $\mathrm{St}(t=0)=3$, 6 and 10 progressively reaching values close to unity, while particles with $\mathrm{St}(t=0)=0.5$ remain below one throughout the decay. The PINN case displays a more complex behaviour:
a short-lived increase in $k_\eta$, which indicates that the range of scales involved in the energy cascade is increasing, 
produces a transient rise of the instantaneous Stokes number before the subsequent decay drives all cases towards smaller values. Consequently, the particles experience markedly different temporal histories of the same control parameter despite having identical physical properties.

This continuous evolution provides an ideal framework to investigate inertial-particle dynamics under genuinely non-stationary conditions. In contrast to statistically stationary turbulence, where the averaged Stokes number is a fixed parameter characterising a given particle population, the present configuration naturally transforms it into a time-dependent trajectory whose evolution depends on the decay history of the carrier flow. As a result, each simulation explores a broad range of instantaneous Stokes numbers while following a distinct temporal path. Moreover, individual realisations span ranges of $\mathrm{St}$ associated with different clustering mechanisms. This unique feature allows us to assess whether particle clustering is determined solely by the instantaneous value of $\mathrm{St}$ or whether it also depends on the previous evolution of the turbulent flow. As will be shown in the following sections, the latter plays a significant role, with particle clustering retaining a clear dependence on the flow and particle history.

\section{Time evolution of particle clusters evolving from a steady state}\label{sec:steady}

\begin{figure}
\centering
\includegraphics[width=\textwidth]{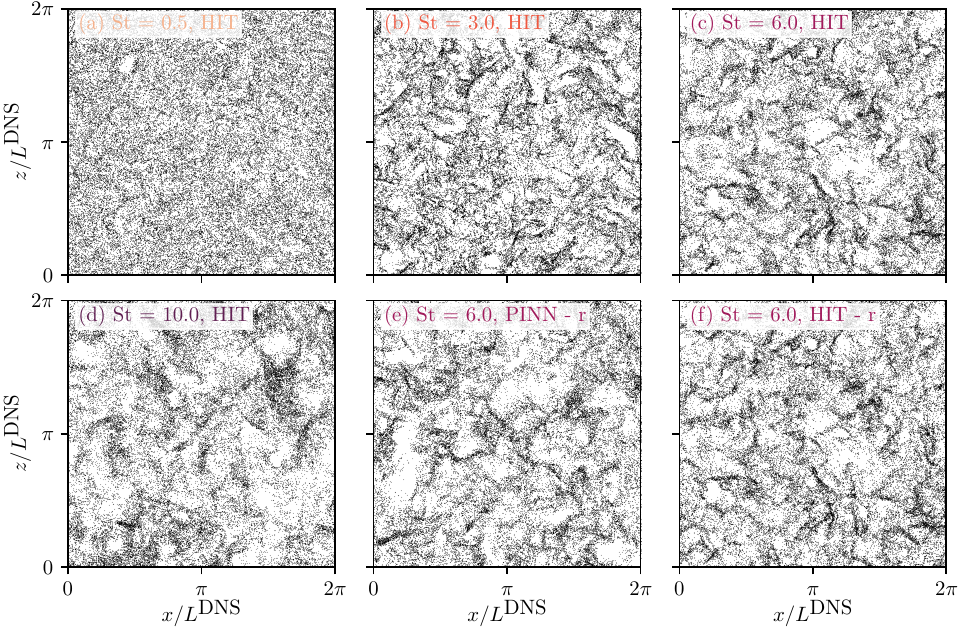}
\hfill
  \caption{Instantaneous visualisation of particles with $y$-coordinate $y/L^\mathrm{DNS}\in[\pi-\Delta, \pi+\Delta]$, with $\Delta = 0.025$, at time $t/T_0 \approx 20$, corresponding to a time within the self-similar decay of the flow. Panels (a-d) correspond to particles in HIT with initial Stokes numbers of $0.5,\,3,\,6$ and $10$, respectively, for which clusters have already formed at $t=0$. Panels (e) and (f) correspond to $\mathrm{St}(t=0) = 6$, evolving in PINN and HIT flows, respectively, from an initial random distribution (labelled by the suffix ``r''). The corresponding movies are provided as supplementary material.}
\label{fig:snapshots}
\end{figure}

In this section we focus on the reference case of pre-clustered particles initially evolving under statistically stationary homogeneous isotropic turbulence before the forcing is removed. The influence of alternative turbulent initial conditions is addressed quantitatively in the following section, although some qualitative comparisons are introduced here to facilitate the discussion of the particle dynamics. As detailed in the previous section, we consider particles with initial Stokes numbers $\mathrm{St}(t=0)=0.5,\,3,\,6$ and $10$, where $t=0$ denotes the instant at which the external forcing is switched off. Since the turbulent kinetic energy decays after this time, the Kolmogorov time scale continuously increases and the particles therefore experience a time-dependent Stokes number throughout the decay.

Figure~\ref{fig:snapshots} presents instantaneous snapshots of the particle distributions for the different initial Stokes numbers and flow configurations, at a time within the self-similar decay of the flow. While the subsequent analysis in this section focuses on the reference HIT case, the snapshots already reveal qualitative differences associated with the initial turbulent organisation. The asymmetric forcing modifies not only the visual appearance of the particle field, but also the evolution of coherent clusters. These differences persist throughout the decay, suggesting that the initial properties of the turbulent flow continue to influence inertial-particle dynamics well after the forcing has been removed. A more extended discussion is presented in the next section.

\begin{figure}
\centering
\includegraphics[width=\textwidth]{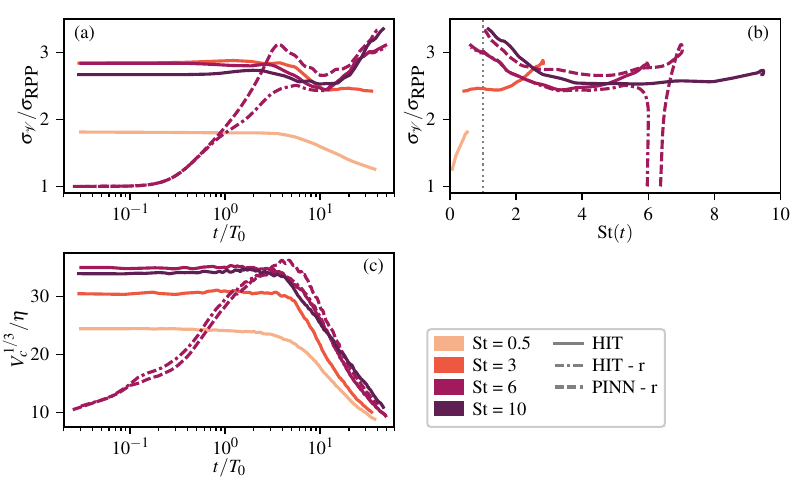}
\hfill
  \caption{Evolution of properties of particle clusters. (a) Standard deviation of Vorono\"i volumes, normalised by the value expected from a random particle distribution, as a function of time. (b) Same but as a function of the time-dependent Stokes number. The vertical dashed line corresponds to $\mathrm{St} = 1$. (c) Linear cluster size in units of the Kolmogorov scale. The labels for all panels are indicated at the top. The colours indicate the initial Stokes number, while the line style indicates the type of flow (HIT or PINN) and the initial condition of the particles at $t=0$ (either from a steady state or from a random distribution, ``r'').}
\label{fig:sigma_voro}
\end{figure}

A quantitative characterisation of these observations is provided by the Vorono\"i analysis shown in figure~\ref{fig:sigma_voro}. The Vorono\"i tessellation partitions the flow domain into cells surrounding each particle such that every point within a given cell is closer to its associated particle than to any other. The normalised cell volumes therefore provide a local measure of particle concentration: small cells correspond to densely populated regions (associated with clusters), whereas large cells identify particle-depleted regions (associated with voids). Following \citet{Monchaux2010}, the standard deviation of the normalised Vorono\"i volumes, $\sigma_{\mathcal{V}}$, is employed as a global measure of clustering intensity, increasing as the particle distribution departs from complete spatial randomness. For a random three-dimensional Poisson distribution (RPP) $\sigma_\mathcal{V} \equiv \sigma_\text{RPP}\approx0.42$~\citep{Tanemura2003, Uhlmann2020}; values exceeding this threshold therefore indicate preferential concentration, with progressively larger values corresponding to increasingly heterogeneous particle distributions. 

Beyond this global metric, the Vorono\"i analysis also enables the identification of coherent particle clusters, from which statistics of their characteristic volumes and spatial organisation can be extracted. A cluster is comprised of at least two adjacent Vorono\"i cells with volume $\mathcal{V}<\mathcal{V}_c$, where $\mathcal{V}_c$ corresponds to the first crossing of the probability density function of $\mathcal{V}$ with the Vorono\"i volume distribution from a RPP \citep{Monchaux2010}.

For $\mathrm{St}=0.5$, the standard deviation of the normalised Vorono\"i volumes remains systematically different throughout the decay compared to all other datasets, regardless of the type of flow, demonstrating that the clustering dynamics retain a clear memory of the initial turbulent state. Particles with larger inertia exhibit an even richer temporal evolution. Rather than relaxing monotonically towards a homogeneous distribution as the turbulence decays, both the clustering intensity and the characteristic cluster volume display pronounced maxima and minima. This behaviour reflects the continuous increase of the Kolmogorov time scale during the decay, which causes the instantaneous Stokes number to evolve and continuously modifies the coupling between the particles and the turbulent flow. Consequently, the evolution of preferential concentration reflects the continuous adjustment of the particle distribution to the evolving turbulent scales and the associated changes in the mechanisms responsible for clustering. The clustering dynamics are therefore not determined solely by the decay of the turbulent kinetic energy, but reflect the evolving particle–turbulence interaction and retain a measurable dependence on the previous evolution of the system.

\begin{figure}
\centering
\includegraphics[width=\textwidth]{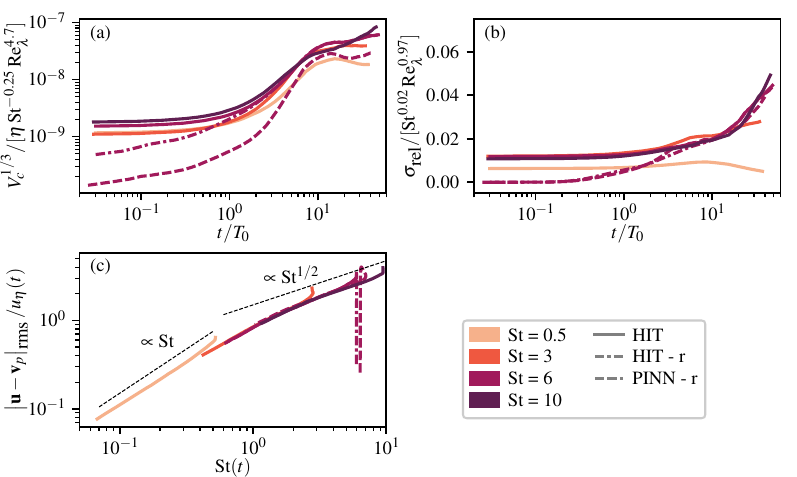}
\hfill
  \caption{
  (a) Linear cluster size and (b) relative Vorono\"i standard deviation $\sigma_\mathrm{rel} = (\sigma_\mathcal{V}-\sigma_\mathrm{RPP})/\sigma_\mathrm{RPP}$, compensated by empirical predictions given by \citet{Sumbekova2017}.
  (c) Particle slip velocity normalised by the instantaneous Kolmogorov velocity, as a function of the instantaneous particle Stokes number. Power-laws with exponents $1$ and $1/2$ are shown as reference.  Labels for all panels are indicated on the bottom right, and are the same as in figure~\ref{fig:sigma_voro}.}
\label{fig:comparison_literature}
\end{figure}

The evolution of the characteristic linear cluster size $V_c^{1/3}$, shown in figure~\ref{fig:sigma_voro}(c), reveals a notable trend. During the decay, the linear cluster size, normalised by the instantaneous Kolmogorov length scale, decreases monotonically for all particle populations. This trend is not obvious a priori. As turbulence decays, all characteristic flow scales increase owing to the progressive reduction of the dissipation rate. One could therefore expect particle clusters to broaden together with the turbulent structures. Instead, when measured relative to the dissipative scale, coherent particle clusters become progressively more compact. This demonstrates that the evolution of the particle field cannot be interpreted as a passive stretching of the initial clustered distribution by the carrier flow.

This observation can be interpreted in the light of previous studies on statistically stationary turbulence. \citet{Sumbekova2017}, using a large experimental dataset of sub-Kolmogorov water droplets in wind tunnels, proposed that the characteristic cluster size follows the empirical scaling
\begin{equation}
\frac{V_c^{1/3}}{\eta}\propto \text{St}^{-0.25}\text{Re}_\lambda^{4.7}.
\end{equation}
Unlike statistically stationary turbulence, however, the present configuration continuously explores the $(\text{St},\text{Re}_\lambda)$ parameter space during a single simulation (figure~\ref{fig:flow_properties}). As shown in figure~\ref{fig:flow_properties}(c)\&(d), both quantities evolve simultaneously but in opposite directions with respect to their influence on clustering. The decrease of the instantaneous Stokes number tends to move particles towards the regime of maximum preferential concentration, whereas the simultaneous reduction of the Reynolds number acts to weaken clustering and reduce the characteristic cluster size. Owing to the much stronger Reynolds-number dependence predicted by the stationary scaling, the latter effect dominates, leading to the systematic decrease of $V_c^{1/3}/\eta$ observed in figure~\ref{fig:sigma_voro}(c).

The comparison with the stationary scaling is further illustrated in figure~\ref{fig:comparison_literature}(a), where the compensated quantity $V_c^{1/3}/(\eta \text{St}^{-0.25} \text{Re}_\lambda^{4.7})$ is shown. A satisfactory collapse is obtained for all cases, indicating that the functional dependence proposed by \citet{Sumbekova2017} remains a useful empirical description of the evolving cluster size in our non-stationary conditions. However, the different particle populations collapse onto distinct prefactors rather than a unique constant. This suggests that, while the instantaneous values of $St$ and $Re_\lambda$ largely govern the evolution of the characteristic cluster size, they do not completely determine it. Instead, the proportionality constant retains a measurable dependence on the previous evolution of the particle field and of the carrier turbulence. In other words, the persistence of the scaling exponents established for statistically stationary turbulence suggests a common underlying self-similar process, while the different prefactors retain the imprint of the system’s previous evolution.

Figure~\ref{fig:comparison_literature}(c) presents the root-mean-square particle slip velocity, normalised by the instantaneous Kolmogorov velocity, as a function of the instantaneous Stokes number. The particle slip velocity is one of the fundamental quantities governing inertial-particle dynamics. Beginning with the centrifuge mechanism proposed by \citet{Maxey1987}, inertial clustering has been interpreted as a consequence of the inability of particles to instantaneously follow the surrounding fluid motion. More recent theoretical developments have generalised this picture by identifying the slip velocity as the quantity controlling the degree of particle--flow coupling and, consequently, preferential sampling, particle accelerations, relative velocities and collision statistics \citep{Bec2003,Gustavsson2016}. In particular, \citet{brandt2022particle} and \citet{Berk2024} argue that the slip velocity provides a unifying framework for understanding inertial-particle dynamics across a broad range of regimes, as it directly quantifies the departure of particle trajectories from those of fluid tracers.

The present results are fully consistent with this interpretation. Two distinct scaling regimes clearly emerge. For $St\lesssim1$, the slip velocity increases approximately linearly with the Stokes number, indicating that particles remain strongly coupled to the dissipative turbulent motions. For $St\gtrsim1$, the dependence weakens towards $|u-v_p|_{\mathrm{rms}}/u_\eta\propto St^{1/2}$, reflecting the progressive decoupling of particles from the smallest turbulent scales. These two asymptotic regimes coincide with those discussed by \citet{brandt2022particle} and demonstrate that, despite the non-stationary evolution of the carrier turbulence, the instantaneous particle--flow coupling remains essentially governed by the instantaneous Stokes number.

The behaviour of the slip velocity also highlights the central result of the present work. If the instantaneous particle--flow coupling were sufficient to determine preferential concentration, the clustering statistics should collapse when represented as a function of the instantaneous Stokes number. However, figure~\ref{fig:sigma_voro}(b) demonstrates that this is not the case. 
Therefore, while the slip velocity retains the classical behaviour expected for statistically stationary turbulence, the spatial organisation of the particles preserves an additional dependence on the previous evolution of both the carrier flow and the particle distribution. The implications of these two coupling regimes for the persistence of preferential concentration and the role of the initial conditions are discussed in the following section.

\section{Effect of initial conditions on the decay of particle clusters}\label{sec:initial}

\begin{figure}
\centering
\includegraphics[width=\textwidth]{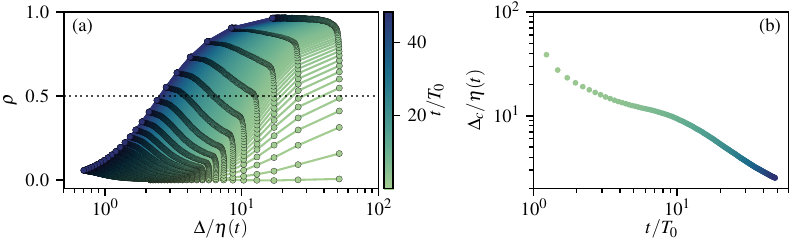}
\hfill
  \caption{
  (a) Pearson correlation coefficient $\rho$ between instantaneous coarse-grained particle distributions in HIT corresponding to $\mathrm{St}=6$ starting from a clustered and a random distribution, as a function of $\Delta$, the size of the coarse-graining scale. The colour bar indicates time in units of $T_0$. (b) Scale $\Delta_c$ where $\rho=\rho_c=0.5$ (indicated by the dotted line in panel a) as a function of time, and normalised by the instantaneous Kolmogorov scale $\eta(t)$.
  }
\label{fig:correlation_HIT}
\end{figure}

We now focus on how the initial conditions of both the flow and the particle distribution affect the  properties of clusters as turbulence decays. We consider particles with initial Stokes number $\text{St}(t=0) = 6$ which are randomly distributed at $t=0$, both in HIT and in the PINN-generated flow. These sets of data are labelled ``HIT - r'' and ``PINN - r'', respectively, where the `r' indicates the initial random distribution of the particles.
As briefly mentioned in the previous sections, comparing panels (e) and (f) of figure~\ref{fig:snapshots} shows that the PINN case displays a distinct spatial distribution compared to HIT, in particular voids appear larger. This difference is reflected in the evolution of $\sigma_\mathcal{V}$ in figure~\ref{fig:sigma_voro}(a)\&(b), with the PINN case having a significantly larger standard deviation for $t>T_0$, consistent with the presence of larger voids, which are primarily responsible for the larger values of $\sigma_\mathcal{V}$.
In spite of these differences, the evolution of the characteristic cluster size shown in figure~\ref{fig:sigma_voro}(c) is similar between HIT and the PINN-generated flow, when measured in units of the time-dependent Kolmogorov scale $\eta(t)$. While clusters reach a slightly larger size for the PINN case at around $t/T_0\approx 5$, the time around when the self-similar decay starts, both throughout the initial stage and the later-decay values are comparable for the two types of flow. 

Interestingly, when we consider again the empirical scaling proposed by
\citet{Sumbekova2017} for the relative Vorono\"i standard deviation,
\begin{equation}
    \sigma_\text{rel}\equiv
    \frac{\sigma_\mathcal{V}-\sigma_\text{RPP}}{\sigma_\text{RPP}}
    \propto \text{St}^{0.02}\text{Re}_\lambda^{0.97},
\end{equation}
as shown in figure~\ref{fig:comparison_literature}(b), the two datasets exhibit
nearly overlapping behaviour. Although the compensated quantity does not
approach a constant value, accounting for the instantaneous Stokes and Reynolds
numbers substantially reduces the differences between the two flow
configurations. This suggests that a significant part of the observed
differences in clustering intensity can be attributed to their distinct
evolutions in parameter space, while the remaining temporal dependence reflects
the non-stationary nature of the problem. Similarly, the slip velocities shown
in figure~\ref{fig:comparison_literature}(c) exhibit comparable behaviour for
the two flow configurations once clustering has developed.

We also evaluate how the initial particle distribution affects the subsequent decay. To this end, we compare particles with $\mathrm{St}(t=0)=6$ in HIT, starting either from an initially clustered state (`HIT') or from an initially random distribution (`HIT - r'). To only assess the effect of the initial conditions of the particles the underlying velocity field is identical in the two cases. A visual comparison between panels (c) and (f) of figure~\ref{fig:snapshots} reveals striking similarities between the two: particles appear to be organised around the same flow structures (this is also observed in the movies provided as supplementary material). We also observe similarities between HIT and HIT-r when $\text{St}(t=0) = 3$.  Even though it can be expected that particles will develop clusters with similar statistical properties (as is the case for $\sigma_\mathcal{V}$ shown in figure~\ref{fig:sigma_voro}(a)), because all known clustering mechanisms are ultimately associated with correlations between the particle distribution and the topology of the carrier flow, the similarities appear to extend beyond statistical measures. Clusters and voids are found in approximately the same regions of the domain, despite the different initial particle distributions. 

To quantify this similarity, we compute a Pearson correlation coefficient $\rho$ between the two sets of data at each time step. We first construct a coarse-grained version of the particle distribution by binning the particles' positions using bins of size $\Delta$. We then compute $\rho$ as
\begin{equation}
    \rho_{xy}(t) = \frac{\sum\limits_i (x_i-\overline{x})(y_i-\overline{y})}{\sqrt{\sum\limits_i (x_i-\overline{x})^2}\sqrt{\sum\limits_i (y_i-\overline{y})^2}},
\end{equation}
where $x_i, y_i$ correspond to the number of particles in the $i$-th bin at time $t$ from each of the datasets considered, and the over-line denotes the mean value over all bins.
Figure~\ref{fig:correlation_HIT}(a) shows $\rho$ as a function of the coarse-graining scale $\Delta$ normalised by the time-dependent Kolmogorov scale, for different times (indicated by the colour bar). 

Shortly after the forcing is removed, the correlation rapidly increases and remains remarkably high throughout the decay. Depending on the coarse-graining scale, the correlation coefficient typically exceeds $\rho\approx0.5$ (indicated by the pointed horizontal line) and reaches values close to $\rho\approx0.9$ at the largest scales. In turbulent flows, correlation levels of this magnitude are generally associated with a very strong spatial correspondence and are commonly used to identify coherent flow structures and dynamically coupled regions of the flow \citep{He2017,Iacobello2018}. Such values indicate that a substantial fraction of the spatial organisation of the particle field is shared by the two realisations, despite their fundamentally different initial conditions. Indeed, as observed in figure~\ref{fig:sigma_voro}(a), for times $t/T_0 \gtrsim 10$ the curves are nearly indistinguishable. This similarity is also reflected in the high values attained by the instantaneous correlation coefficient for a wide range of scales $\Delta$. We identify the critical value $\Delta_c(t)$ such that $\rho(t,\Delta_c) = \rho_c = 0.5$, which is shown in figure~\ref{fig:correlation_HIT}b as a function of time, normalised by $\eta(t)$. Different values of $\rho_c$ in the range $0.3<\rho_c<0.75$ result in similar trends for $\Delta_c$. For times $t/T_0 \gtrsim 10$, the particle distributions become significantly similar (i.e. $\rho \geq \rho_c$) at scales $\Delta_c\approx 10\eta$. This indicates that the spatial correspondence between the two particle distributions is not confined to the largest scales of the flow, but extends down to coarse-graining scales of approximately $10\eta$.

This level of agreement is far from trivial: although both particle populations evolve in the same velocity field and are therefore expected to respond to the same attracting regions of the turbulence, it is not obvious a priori that they should reconstruct nearly the same arrangement of clusters and voids. As particles have a finite response time, they do not adjust instantaneously to the surrounding flow, so they do not necessarily relax to the same positions when they originate from different initial distributions.
The results instead demonstrate that the evolving turbulent field imposes a robust geometrical organisation on the particle distribution, rapidly reducing the influence of the initial particle positions on the large-scale spatial organisation.

It is important to emphasise, however, that this convergence should not be interpreted as a complete loss of memory. Rather, it indicates that the large-scale spatial skeleton of preferential concentration is predominantly determined by the instantaneous turbulent structures. Remarkably, this appears to hold for all the Stokes numbers considered, spanning different degrees of coupling between the flow and particle inertia. In contrast, other statistics, including the clustering intensity (figure~\ref{fig:sigma_voro}b), the characteristic cluster size (figure~\ref{fig:sigma_voro}c), and the scaling prefactors discussed in figure~\ref{fig:comparison_literature}, retain a measurable dependence on the initial conditions. The evolution of inertial particle clusters is therefore characterised by two complementary behaviours: while the large-scale geometry rapidly converges towards a nearly universal configuration imposed by the carrier flow, finer statistical properties preserve a measurable memory of both the initial particle distribution and the preceding evolution of the turbulence.

\section{Conclusions}
\label{sec:discussion}

We have investigated the evolution of inertial-particle clustering during the free decay of homogeneous isotropic turbulence by means of direct numerical simulations. In contrast to statistically stationary turbulence, where the particle Stokes number is fixed, freely decaying turbulence provides a natural framework in which the instantaneous Stokes number evolves continuously while the particle properties remain unchanged. By considering particles initially distributed either randomly or in statistically stationary clustered configurations, together with distinct turbulent initial conditions, we have separated the effects of the instantaneous flow state from those associated with the previous evolution of the particle--flow system.

The principal result is that preferential concentration in decaying turbulence cannot be characterised by the instantaneous Stokes number alone. This conclusion does not imply, however, a general breakdown of relationships established under statistically stationary conditions. Instead, our results identify a clear distinction between different aspects of inertial-particle dynamics. The particle slip velocity follows the classical dependence on the instantaneous Stokes number, recovering the two regimes associated with strongly coupled and progressively decoupled particles. Thus, the instantaneous particle--flow coupling remains well described by the conventional Stokes-number framework even as the turbulence evolves. Clustering statistics, by contrast, do not collapse when expressed solely in terms of the instantaneous Stokes number and retain a measurable dependence on the preceding evolution of both the particles and the carrier flow.

This distinction is further quantified by the comparison with empirical models developed for statistically stationary turbulence. The characteristic cluster size decreases monotonically relative to the instantaneous Kolmogorov scale during the decay, but its evolution remains consistent with the dependence on $\mathrm{St}$ and $\mathrm{Re}_\lambda$ proposed for stationary turbulence. The persistence of the corresponding scaling exponents suggests that the evolution is governed by a similar underlying self-similar process. The different datasets nevertheless remain separated by their proportionality constants, showing that the history of the particle--flow system is encoded in the prefactor rather than in the scaling exponents. Similarly, compensation of the clustering intensity by its stationary dependence on $\mathrm{St}$ and $\mathrm{Re}_\lambda$ substantially reduces the differences between distinct turbulent initial conditions, although it does not remove their temporal evolution. Existing stationary scalings therefore retain predictive value in decaying turbulence, but require an additional history-dependent contribution to provide a complete description.

The comparison between initially clustered and initially random particle populations reveals a complementary and, importantly, spatial manifestation of this behaviour. Despite their markedly different configurations at the onset of decay, populations evolving in the same velocity field rapidly develop strongly correlated spatial distributions. The correlation reaches values close to $\rho\simeq0.9$ at large coarse-graining scales, while for $t/T_0\gtrsim10$ significant correspondence extends down to scales of approximately $10\eta$. The carrier turbulence therefore imposes a robust spatial skeleton on preferential concentration that becomes largely independent of the initial particle positions. This convergence is not merely statistical: clusters and voids generated from different initial particle distributions preferentially occupy approximately the same regions of the flow.

Taken together, these results reveal a separation between the geometry and the statistics of preferential concentration in non-stationary turbulence. The large-scale spatial organisation is rapidly selected by the evolving carrier flow, whereas clustering intensity, characteristic cluster size and their scaling prefactors retain information about the previous evolution of the particle--flow system. At the same time, quantities directly characterising the instantaneous particle--flow coupling, such as the slip velocity, remain primarily controlled by the instantaneous Stokes number. History dependence is therefore not a uniform correction to inertial-particle dynamics, but affects different observables in fundamentally different ways.

These findings place a specific constraint on the use of models derived from statistically stationary turbulence in transient flows. Relationships based on instantaneous $\mathrm{St}$ and $\mathrm{Re}_\lambda$ can remain applicable to the local particle--flow coupling and can capture the functional dependence of several clustering quantities, but they cannot in general determine preferential concentration uniquely. A complete description requires information about the preceding particle and flow dynamics, for instance through history-dependent amplitudes or additional state variables. This distinction should be particularly relevant to natural and industrial particle-laden flows in which turbulence evolves on time scales comparable to the particle response and particles enter transient or decaying regions with an already established spatial organisation.


\appendix
\section{Protocol for generating flow initial conditions}
\label{sec:methods}
To generate flow initial conditions that are compatible with HIT while exhibiting systematic departures in third-order statistics, we employ a protocol combining physics-informed neural networks (PINNs) with the data-assimilation technique of nudging. Full details of the validation of this methodology, together with an analysis of the decay properties of the resulting fields, are provided by \citet{Angriman2023,Angriman2024}. We implement a neural network whose output is constrained to follow the incompressible Navier--Stokes equations while reproducing a prescribed central third-order moment. To this end, we define the loss function
\begin{equation}
    L = L_d + \lambda_p L_p + \lambda_s L_s,
    \label{eq:loss_function}
\end{equation}
where
\begin{equation}
    L_d = \frac{1}{N_b}\sum_{\{i\}} \left({\bm u}_i-{\bm u}_i^0\right)^2
\end{equation}
is the data term and ${\bm u}_i^0$ denotes the initial seed. The index $i$ identifies the spatial location and time at which the fields are evaluated, i.e. ${\bm u}_i={\bm u}(x_i,y_i,z_i,t_i)$, and the sum is taken over $N_b$ samples in the mini-batch $\{i\}$. The hyperparameters $\lambda_p$ and $\lambda_s$ control the relative contributions of the different terms to the total loss function. The physics term,
\begin{equation}
    L_p =
    \frac{1}{N_b}
    \sum_{\{i\}}
    \left[
    \left(
    \frac{\partial {\bm u}_i}{\partial t}
    +({\bm u}_i\cdot{\bm \nabla}){\bm u}_i
    +{\bm \nabla}p_i
    -\nu\nabla^2{\bm u}_i
    \right)^2
    +
    \left({\bm \nabla}\cdot{\bm u}_i\right)^2
    \right],
\end{equation}
constrains ${\bm u}$ to satisfy the incompressible Navier--Stokes equations, with $p$ denoting the pressure divided by the fluid density. The statistical term,
\begin{equation}
\begin{split}
L_s =
\left(
\frac{1}{N_b}\sum_{\{i\}}u_i
\right)^2
&+
\left[
\sqrt{
\frac{1}{N_b}\sum_{\{i\}}u_i^2
-
\left(
\frac{1}{N_b}\sum_{\{i\}}u_i
\right)^2
}
-\sigma_0
\right]^2
\\
&+
\left[
\frac{1}{N_b}\sum_{\{i\}}
\left(
u_i-\frac{1}{N_b}\sum_{\{i\}}u_i
\right)^3
-s_0^3
\right]^2,
\end{split}
\end{equation}
imposes prescribed statistical moments on the $x$-component of the velocity field, $u$. The first term enforces a zero mean, the second constrains the standard deviation to $\sigma_0$, and the third imposes a prescribed central third-order moment $s_0^3$. We set $\sigma_0=0.5\,U^\mathrm{DNS}$, such that the characteristic velocity remains of order unity in the adopted nondimensionalisation, and $s_0=0.25\,U_0$. Further details of the choice of hyperparameters and of the training procedure are given in \citet{Angriman2023}.

The initial seed ${\bm u}^0$ is sampled from a low-resolution HIT simulation with $32^3$ grid points. The gradient of equation~\eqref{eq:loss_function} is then evaluated iteratively and the neural-network weights are updated until the prescribed statistical moments converge within fluctuations. The resulting low-resolution PINN field is subsequently interpolated onto the target resolution, $512^3$ grid points in the present case. To ensure that the small scales of the reconstructed field are dynamically compatible with the Navier--Stokes equations, we then apply nudging. Specifically, we solve equations~\eqref{eq:NS} with the additional forcing term
\begin{equation}
    {\bm f}_{\mathrm{nudg}}
    =
    -\alpha\,\mathcal{I}
    \left(
    {\bm u}-{\bm u}_{\mathrm{ref}}
    \right),
\end{equation}
on the right-hand side, where ${\bm u}_{\mathrm{ref}}$ is the reference field obtained from the final PINN state. This term penalises departures of the evolving velocity field from the reference field. The parameter $\alpha$ controls the strength of the nudging, while $\mathcal{I}$ denotes a low-pass filter in Fourier space that restricts the penalisation to the scales at which reference data are prescribed. In the present case, the filter acts on the Fourier shell $kL^\mathrm{DNS}\in[0,9]$.

After application of the combined PINN--nudging protocol, the resulting velocity field exhibits a broad inertial range consistent with turbulent dynamics while retaining, at large scales, the statistical properties imposed through the PINN reconstruction. In particular, the final field remains consistent with the prescribed low-order and third-order moments of the reference state.

\backsection[Declaration of interests]{The authors report no conflict of interest.}

\bibliographystyle{jfm}
\bibliography{biblio}

\end{document}